\documentclass[reprint,superscriptaddress,amsmath,amssymb,prb,longbibliography]{revtex4-1}
\usepackage{comment}
\usepackage{color}
\usepackage{graphicx}
\usepackage{subfigure}
\usepackage{dcolumn}
\usepackage{bm}
\usepackage{amsmath}
 
\usepackage{braket}
\usepackage{diagbox, eqparbox, hhline}
\usepackage[detect-none]{siunitx}
\DeclareMathAlphabet{\mathpzc}{OT1}{pzc}{m}{it}

\begin{document}

\title{Spin textures in chiral magnetic monolayers \\with suppressed nearest-neighbor exchange}

\author{Ra\'i M. Menezes}
\affiliation{Department of Physics, University of Antwerp, Groenenborgerlaan 171, B-2020 Antwerp, Belgium}
\affiliation{NANOlab Center of Excellence, University of Antwerp, Belgium}
\affiliation{Departamento de F\'isica, Universidade Federal de Pernambuco, Cidade Universit\'aria, 50670-901, Recife-PE, Brazil}

\author{Cl\'ecio C. de Souza Silva}
\affiliation{Departamento de F\'isica, Universidade Federal de Pernambuco, Cidade Universit\'aria, 50670-901, Recife-PE, Brazil}

\author{Milorad V. Milo\v{s}evi\'c}
\email{milorad.milosevic@uantwerpen.be}
\affiliation{Department of Physics, University of Antwerp, Groenenborgerlaan 171, B-2020 Antwerp, Belgium}
\affiliation{NANOlab Center of Excellence, University of Antwerp, Belgium}

\begin{abstract}
    High tunability of two dimensional magnetic materials (by strain, gating, heterostructuring or otherwise) provides unique conditions for studying versatile magnetic properties and controlling emergent magnetic phases. Expanding the scope of achievable magnetic phenomena in such materials is important for both fundamental and technological advances. Here we perform atomistic spin-dynamics simulations to explore the (chiral) magnetic phases of atomic monolayers in the limit of suppressed first-neighbors exchange interaction. We report the rich phase diagram of exotic magnetic configurations, obtained for both square and honeycomb lattice symmetries, comprising coexistence of ferromagnetic and antiferromagnetic spin-cycloids, as well as multiple types of magnetic skyrmions. We perform a minimum-energy path analysis for the skyrmion collapse to evaluate the stability of such topological objects, and reveal that magnetic monolayers could be good candidates to host the antiferromagnetic skyrmions that are experimentally evasive to date. 
\end{abstract}

\date{\today}
\pacs{Valid PACS appear here}
\maketitle

\section{Introduction}

Magnetism in two dimensions (2D) has recently drawn immense attention of both theoretical and experimental research, due to its fundamental significance and promising technological applications\cite{gibertini2019magnetic,burch2018magnetism}. Magnetic 2D atomic crystals present a diapason of possibilities for controlling magnetic interactions by different composition and structural arrangements\cite{ferriani2007magnetic,hardrat2009complex,meyer2017dzyaloshinskii,xu2020topological}, as well as engineering techniques\cite{huang2018electrical,webster2018strain}. The competition between the magnetic interactions in such 2D materials, e.g., crystalline anisotropy, exchange, dipole-dipole, Dzyaloshinskii-Moriya interaction\cite{dzyaloshinsky1958thermodynamic,moriya1960anisotropic} (DMI), etc., is likely to lead to a wide range of physical phenomena and novel magnetic phases.             
It is well known that the exchange interaction between neighboring spins of a magnetic system plays a determinant role in resulting magnetic configurations. Interestingly, the exchange stiffness of magnetic monolayers can be tuned in a multitude of ways. For example, a manganese monolayer presents ferromagnetic (FM) order when grown on the (001) surface of tungsten substrate\cite{ferriani2008atomic}, but presents antiferromagnetic (AFM) order when grown on tungsten (110)\cite{sessi2009temperature}. Ref.~\onlinecite{ferriani2007magnetic} claimed based on first-principles calculations that the nearest-neighbor (NN) exchange interaction of an iron monolayer on the (001) surface of a Ta$_x$W$_{1-x}$ alloy can be continuously tuned from FM to AFM coupling by varying the Ta concentration in the substrate, and that in the situation of weak NN exchange nontrivial magnetic configurations can be achieved. Similarly, Ref.~\onlinecite{tan2019tunable} showed that the ground state of a Fe monolayer on top of different 4d and 5d nonmagnetic metals can be continuously manipulated from FM to AFM by in-plane biaxial and uniaxial strain on the substrates. Ref.~\onlinecite{meyer2017dzyaloshinskii} demonstrated tunability of the exchange interaction by changing the stacking order of Fe/5d bilayers on Rh(001). The exchange interaction can also be tuned by strain in magnetic 2D monolayer chromium trihalides CrX$_3$ (with X = I, Cl and Br), continuously from FM to AFM one\cite{webster2018strain}. 

In addition to this (symmetric) exchange interaction, the asymmetric Dzyaloshinskii-Moriya interaction plays an important role in the ordering of a magnetic system. Instead of (anti)parallel spins, DMI favors the rotation of magnetization at short length scales, giving rise to chiral spin structures, such as cycloids and magnetic skyrmions. DMI can be tuned in 2D magnets by breaking inversion symmetry, e.g., in Janus structures of chromium trihalides~\cite{xu2020topological} and manganese dichalcogenides\cite{liang2019very}, or at the interface of the magnetic monolayer with a heavy-metal substrate\cite{bogdanov2001chiral}.

In spite of the many possibilities for tuning exchange interactions in 2D magnets, the resultant magnetic states of such systems remain scarcely investigated. Therefore, in this work we explore the magnetic phases of chiral magnetic monolayers, such as Fe monolayer and CrX$_3$ lattices, in the limit of suppressed NN exchange interaction. We present the rich phase diagram obtained for both square and honeycomb symmetries, where exotic magnetic configurations can be stabilized. There we reveal states with coexisting FM and AFM spin-cycloids and magnetic skyrmions, as well as the novel $p$-AFM skyrmion state, and explore stability of magnetic skyrmions when dominated either by NN or second-nearest-neighbor (SNN) exchange coupling. 

The paper is organized as follows. In Sec.~\ref{s2} we introduce the atomistic spin model used to describe the chiral magnetic monolayer. Sec.~\ref{s3} contains all our results and discussions, starting from the effects of NN and SNN exchange coupling on the magnetic ground state in the presence of dipole-dipole interactions (Sec.~\ref{s3.1}). In Sec.~\ref{s3.2} we present the phase diagram for varied DMI and anisotropy interactions in the limit of suppressed NN exchange in a square lattice, and derive analytical expressions for the phase boundaries in Sec.~\ref{s3.3}. In Sec.~\ref{s3.4} we discuss several types of magnetic skyrmions stabilized in the considered system, and perform minimum-energy path calculations to evaluate the stability of magnetic skyrmions when dominated either by NN or SNN exchange coupling. Sec.~\ref{ss3C} contains the phase diagram and notes for a honeycomb lattice, comparatively to the discussion presented for a square lattice in Sec.~\ref{s3.2}. Our findings are summarized in Sec.~\ref{s5}.

\section{Atomistic spin model}\label{s2}
In this work we perform atomistic spin simulations to capture possible magnetic phases of chiral magnetic monolayers, primarily based on the simulation package \textit{Spirit} (see Ref. \onlinecite{muller2019spirit}). The extended Heisenberg Hamiltonian of the considered classical system of spins, in absence of applied magnetic field, is given by
\begin{eqnarray}
    \mathcal{H}=&-&\sum_{<i,j>_\text{sd}}J_{ij}\textbf{n}_i\cdot\textbf{n}_j-\sum_{<i,j>_\text{st}}\textbf{D}_{ij}\cdot(\textbf{n}_i\times\textbf{n}_j)\nonumber\\
    &-&\frac{1}{2}\mathcal{D}_\text{ddi}\sum_{i,j\neq i}\frac{3(\textbf{n}_i\cdot\bm{\hat{r}}_{ij})(\textbf{n}_j\cdot\bm{\hat{r}}_{ij})-(\textbf{n}_i\cdot\textbf{n}_j)}{(r_{ij}/r_0)^3}\nonumber\\
    &-&K\sum_i(\textbf{n}_i\cdot\bm{\hat{z}})^2,
\label{eq.H}
\end{eqnarray}
where $\bm{\mu}_i$ is the magnetic moment of the $i^{th}$ atomic site, with moment $|\bm{\mu}_i|=\mu$ and $\textbf{n}_i=\bm{\mu}_i/\mu$ the $i^{th}$ spin orientation. $J_{ij}$ is the exchange stiffness, where we define $J_1$ and $J_2$ as the NN and SNN exchange stiffness respectively; $\textbf{D}_{ij}=D(\bm{\hat{r}}_{ij}\times\bm{\hat{z}})$ is the DMI vector, with $D$ the DMI strength and $\bm{r}_{ij}$ the vector connecting spins $i$ and $j$. $K$ is the perpendicular magnetic anisotropy
and $\mathcal{D}_\text{ddi}=\mu_0\mu^2/(4\pi r_0^3)$ defines the magnitude of the dipole-dipole interaction (DDI), with $r_0$ the nearest-neighbor distance and $\mu_0$ the vacuum permeability. $< >_\text{st}$ and $<>_\text{sd}$  denote summation up to first and second-neighbors sites respectively. For the dipole-dipole interaction we make use of fast Fourier transforms and the convolution theorem \cite{hayashi1996calculation} adapted to treat arbitrary spin lattice configurations, as implemented in \textit{Spirit}, which reduces significantly the computational effort. Moreover, a direct summation of dipoles has been performed in order to verify selected results. The energy minimization is performed using a Verlet-like velocity projection method, as explained in Refs.~\onlinecite{muller2019spirit,bessarab2015method}, which accelerates convergence towards local minima and avoids overstepping due to momentum considered in the standard Landau-Lifshitz-Gilbert equation~\cite{landau1992theory,gilbert2004phenomenological}. In this method the spins are treated as massive particles moving on the surfaces of spheres, where the velocity at each time step is damped by projecting it along the force $\textbf{F}_i=\partial\mathcal{H}/\partial\textbf{n}_i$.

When constructing the equilibrium phase diagram of a spin system in the space of two relevant parameters in Eq. \eqref{eq.H}, the spin-relaxation simulations are performed for a uniform $10\times10$ matrix of values within the parametric range of the phase diagram, each initialized from a random configuration at least 10 times. After the energy is minimized numerically for those values of parameters, all obtained states are followed ``adiabatically'' for gradually varied parameters, so that ground-state phases could be reliably determined on a high-density grid in the parametric space (typically a $200\times200$ grid for the phase diagrams shown in the next section).

\section{Results and discussion}\label{s3}

\subsection{Competing exchange interactions}\label{s3.1}
\begin{figure}[t]
\centering
\includegraphics[width=0.7\linewidth]{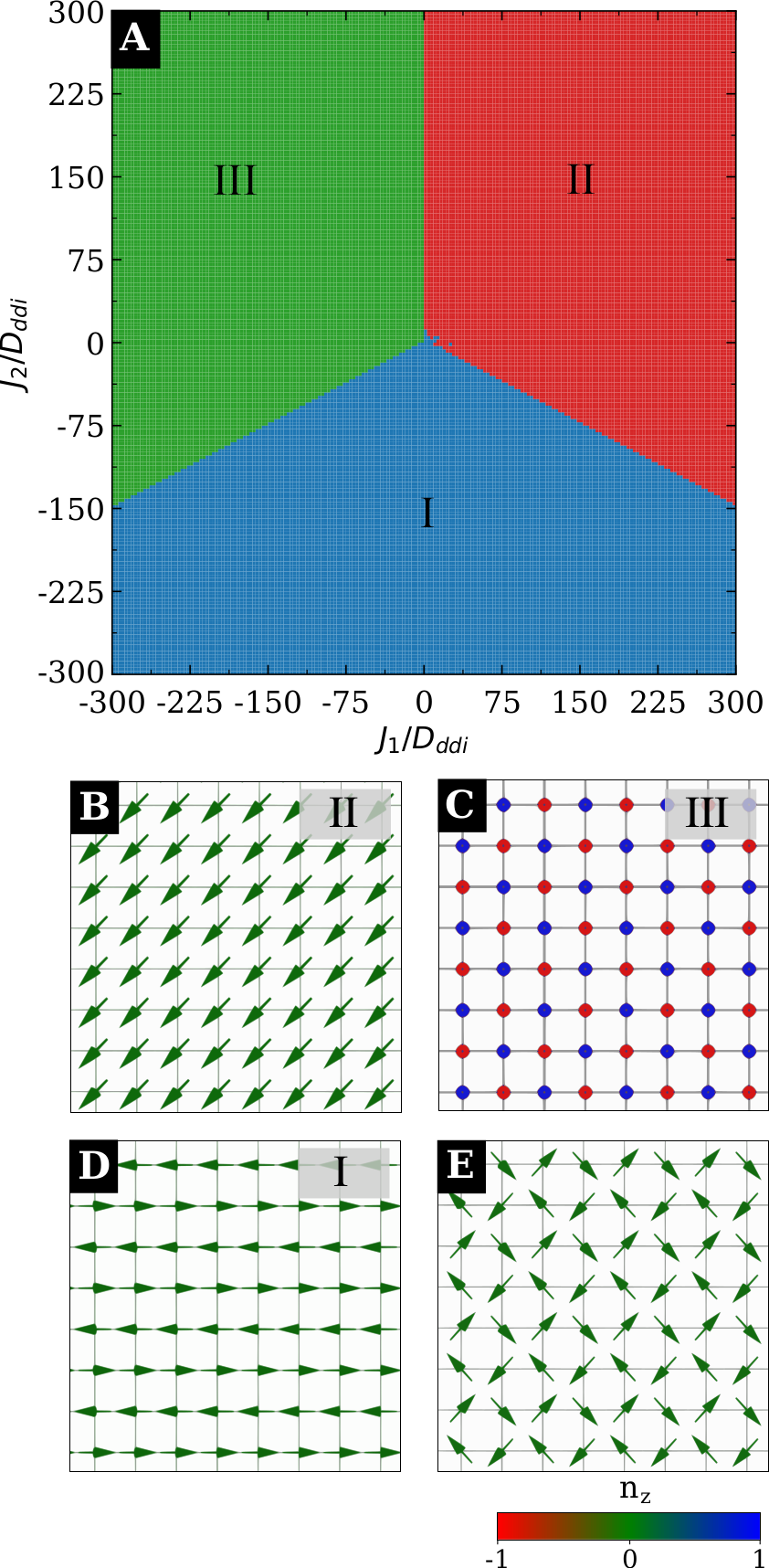}
\caption{(A) Phase diagram for a square monolayer lattice for different values of exchange stiffness $J_1$ and $J_2$, in the presence of dipole-dipole interactions, for $K=D=0$. (B-D) Ground-state magnetic phases, corresponding to nomenclature indicated in (A). Note that in case of dominating exchange interaction the non-collinear and collinear states become nearly energetically degenerate within phase I, so that states similar to the one shown in panel (E) frequently appear as stable.}
\label{fig1}
\end{figure}
In this work we are interested in magnetic configurations that emerge in the limit of vanishing nearest-neighbor exchange interaction. In order to be able to systematically discern effects from different types of interactions, we remove DMI and anisotropy and examine purely the effects of tuning the exchange coupling between both nearest-neighbor and second-nearest-neighbor sites, with dipole-dipole interactions taken into account. 

\subsubsection{Square lattice}
We first consider a square lattice of spin sites, representative of e.g. Fe monolayer on a substrate~\cite{ferriani2007magnetic}. Fig.~\ref{fig1}(A) shows the ground-state phase diagram obtained in the numerical experiments based on Eq.~\eqref{eq.H}. The corresponding minimal energy configurations belonging to different regions of the phase diagram in Fig.~\ref{fig1}(A) are depicted in Fig.~\ref{fig1}(B-D). For dominating nearest-neighbor exchange interaction ($|J_1|>2|J_2|$), only two states can occur, the FM state [Fig.~\ref{fig1}(B)] for $J_1>0$, and the checkerboard $c(2$x$2)$-AFM state [Fig.~\ref{fig1}(C)], also referred to as $c$-AFM, for $J_1<0$. Notice that dipole-dipole interaction favors in-plane spin configurations except for the $c$-AFM phase, where the ``head-to-head" arrangement of spins is not energetically favored, forcing the system to align out of plane. However, if $J_2<-|J_1|/2$ the exchange term in Eq.~\eqref{eq.H} favors the so-called $p(2$x$1)$-AFM ordering [see Fig.~\ref{fig1}(D)], also referred to as $p$-AFM. The $p$-AFM state is a degenerate solution of the Heisenberg Hamiltonian with respect to rotation of consecutive spins by an angle of $\pm\Omega$, with exchange energy $\mathcal{E}_\text{ex}=4J_2$ per spin. Ferriani \textit{et al.}\cite{ferriani2007magnetic} demonstrated that such degeneracy can be lifted by higher-order interactions such as four-spin and biquadratic ones, where noncollinear states, i.e. with $\Omega\neq 0$, are favored. Here we observed that the presence of dipole-dipole interactions favors collinear configurations. However, for dominating exchange energy, even small fluctuations of the spin configurations can overcome the dipolar ineractions so that noncollinear domains can coexist with the collinear ones. In such conditions, even though the state shown in Fig. \ref{fig1}(C) remains the ground state of the system, the states similar to the one shown in Fig.~\ref{fig1}(E) become increasingly stable and frequently appearing.

\begin{figure}[t]
\centering
\includegraphics[width=0.65\linewidth]{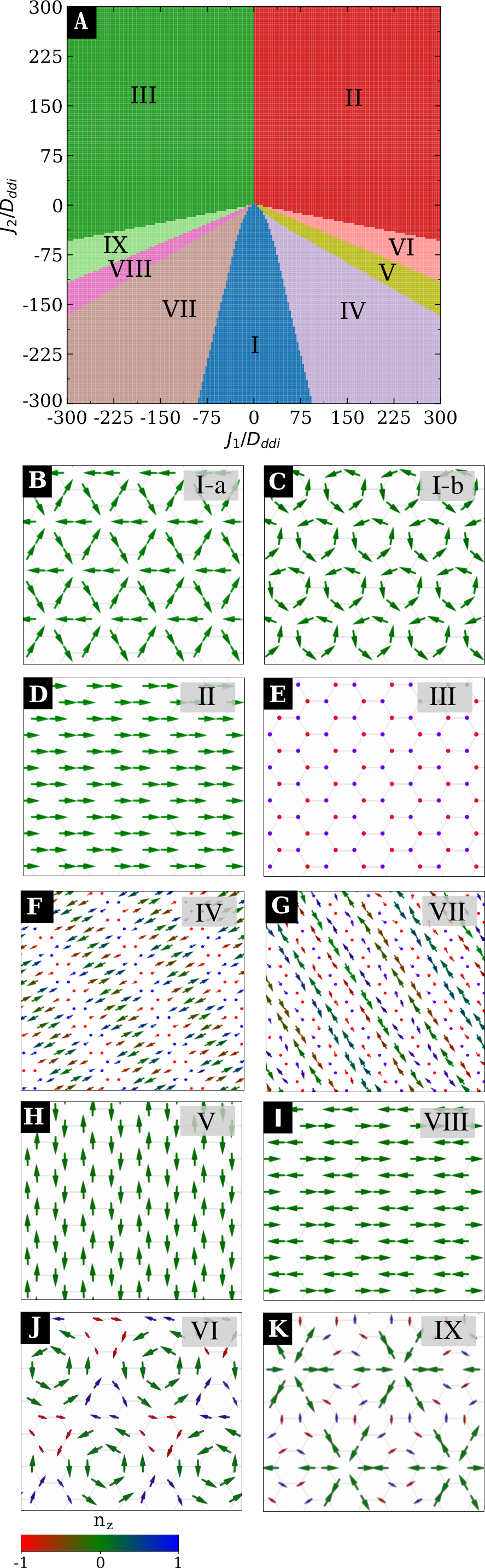}
\caption{(A) Phase diagram for a honeycomb monolayer lattice for different values of exchange stiffness $J_1$ and $J_2$, in the presence of dipole-dipole interactions, for $K=D=0$. (B-K) Ground-state magnetic phases, corresponding to labeling indicated in panel (A).}
\label{fig2}
\end{figure}

\begin{figure*}[t]
\centering
\includegraphics[width=\linewidth]{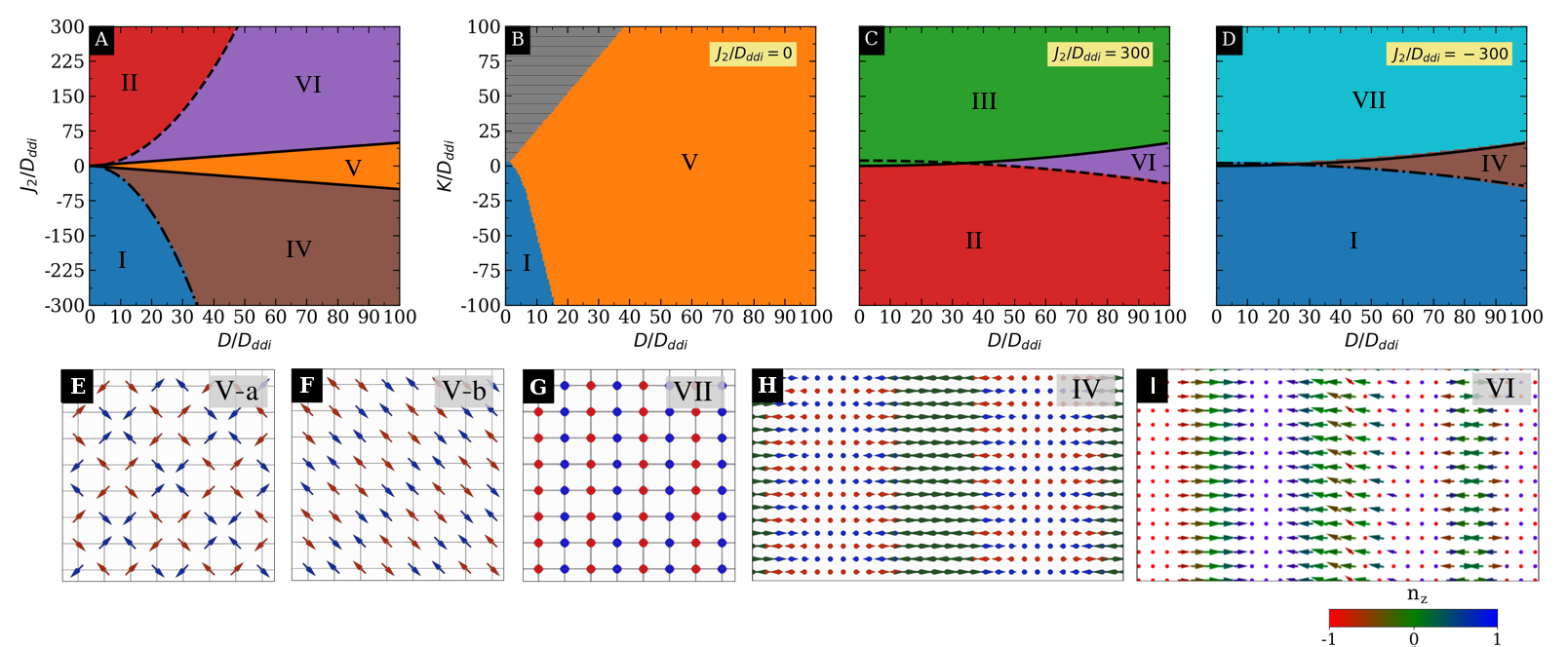}
\caption{Phase diagram for a square monolayer lattice for vanishing nearest-neighbor exchange interaction. (A) As a function of the second-nearest-neighbor exchange $J_2$ and the DMI strength $D$, for $K=0$. (B-D) As a function of magnetic anisotropy $K$ and DMI strength $D$, for $J_2=0$ (B), $J_2=300D_\text{ddi}$ (C) and $J_2=-300D_\text{ddi}$ (D). The patterned region in (B) indicates disordered configurations. (E-I) Spin configurations corresponding to phases indicated in (A-D). Solid lines in panels (A), (C, (D) represent analytical solutions for the phase boundaries of the cycloidal states derived in Sec.~\ref{s3.3}.}
\label{fig3}
\end{figure*}

\subsubsection{Honeycomb lattice}
In Fig.~\ref{fig2}(A) we show the phase diagram for the honeycomb lattice symmetry, analogous to the one of Fig.~\ref{fig1}(A) obtained for the square lattice. In the honeycomb case, the non-trivial structure, as well as the mismatch in the number of NN and SNN bonds of the honeycomb lattice give rise to several more magnetic phases. Furthermore, one should note that the SNN bonds form a triangular lattice, which is intrinsically frustrated under AFM coupling, which can lead to non-trivial spin textures. For the case of $J_2>0$, the FM and AFM states (phases II and III in Fig.~\ref{fig2}) are analogous to the ones observed in the square lattice. On the other hand, for dominating $J_2<0$ two degenerate lattices of spin-loops are obtained, shown in Fig.~\ref{fig2} (B) and (C). Such magnetic phases are similar to those found for multiferroic hexagonal compounds\cite{fabreges2009spin,tan2016pressure}, which in that case are associated to the $\Gamma_1$ and $\Gamma_3$ irreducible representations of the P6$_3$cm space group\cite{wang2013room}. In our system the spin-loop configurations are also favored by the dipole-dipole interactions. By increasing the magnitude of $|J_1|$, the competing NN and SNN coupling give rise to spin cycloids (phases IV and VII in Fig.~\ref{fig2}), collinear AFM states (phases V and VIII in Fig.~\ref{fig2}) and vortex-like configurations (lattice of FM vortices in phase VI and a lattice of AFM antivortices in phase IX, see Fig.~\ref{fig2}). We notice that for $J_1>0$ the spins tend to form pairs of rows with same orientation [see e.g. Fig.~\ref{fig2}(H)] due to the FM NN coupling and the cycling of the spins in the cycloidal phase [Fig.~\ref{fig2}(F)] is of the Néel type. In contrast, for $J_1<0$, the spins are aligned in single rows and the cycloids are of the Bloch type.

\subsection{Suppressed nearest-neighbor exchange in the square lattice}\label{s3.2}

Having understood the competing effects between the nearest-neighbor and second-nearest-neighbor exchange and dipolar interactions, we proceed to examine spin configurations in the limit of vanishing NN exchange interaction. We start with the square lattice, which, due to its simpler geometry, allows for a more illuminating analysis of cycloids and skyrmions induced by suppressed NN exchange. Moreover, as will be shown, the square lattice encompasses all the main ingredients behind the physics of the chiral magnetic textures in the 2D limit. Supplementary notes on the honeycomb lattice case are given in subsection \ref{ss3C}.

In Fig.~\ref{fig3}(A) we show the phase diagram for the square lattice, obtained for $J_1=0$, as a function of $J_2$ but also as a function of increasing DMI. For negative values of $J_2$, increasing DMI strength favors the formation of $p$-AFM spin-cycloids, shown in Fig.~\ref{fig3}(H), while for positive $J_2$ we report the coexistence of FM and $c$-AFM spin-cycloids with \textit{opposite chiralities} [Fig.~\ref{fig3}(I)]. Notice that when neglecting the DDI term in the Hamiltonian [Eq.~\eqref{eq.H}], the FM and $c$-AFM states are energetically degenerate for the case of $J_1=0$ and $J_2>0$ (see also supplemental figure Fig.~S1(A) for a better visualization of the coexisting FM and c-AFM domains). For the case of dominating DMI energy, all spins are forced orthogonal to each other, driving a transition to an emergent spin-ice type of state (V-a) or a degenerated striped state (V-b), shown respectively in Fig.~\ref{fig3}(E) and Fig.~\ref{fig3}(F). Notice that state V-a presents a checkerboard symmetry with four spins pointing in or four spins pointing out of a same tetragon, in such a way that the local magnetization field is not divergence-free, giving rise to so-called magnetic monopoles~\cite{castelnovo2008magnetic,brooks2014magnetic}. The effects of anisotropy interaction (parameter $K$ in the Hamiltonian) are shown in Fig.~\ref{fig3}(B-D) for $J_2=0$, $300D_\text{ddi}$ and $-300D_\text{ddi}$ respectively. For $J_2=0$ the in-plane order of state V is preserved and the $z$ component of the spins are continuously deformed by anisotropy interaction. For $J_2>0$, the out-of-plane (in-plane) anisotropy favors the $c$-AFM (FM) state, while for $J_2<0$ the out-of-plane anisotropy gives rise to the out-of-plane $p$-AFM state, shown in Fig.~\ref{fig3}(G). 

\subsubsection{Phase boundaries}\label{s3.3}

As will be shown in this section, analytical expressions can be derived for most of the phase boundaries in Fig.~\ref{fig3}. In that derivation we consider the extended Heisenberg Hamiltonian in Eq.~\eqref{eq.H} with suppressed NN exchange interaction, and initially neglect the dipole-dipole interaction.
For a simplified analysis, we assume the cycloidal phases are characterized by uniform rotation of the spins along one easy axis, say $x$, that is
\begin{eqnarray}
    {\rm FM}\quad&& \theta_{m,n} = \frac{2\pi m}{N_p}, \\
    c\text{
    -AFM}\quad&& \theta_{m,n} = -\frac{2\pi m}{N_p} - \pi[1-(-1)^{m+n}], \\
    p\text{
    -AFM}\quad&& \theta_{m,n} = \frac{2\pi m}{N_p} - \pi[1-(-1)^n],
\end{eqnarray}
where index $m$ ($n$) counts spin rows parallel (perpendicular) to the $x$ axis and $N_p$ is the number of spins involved in one period of the cycloid, which is to be determined after minimization of the free energy.

The DMI contribution to the energy of the system can easily be calculated by noticing that in all the above described cycloidal configurations only spins linked along the $x$ direction have nonzero DMI energy. In this case ${\bf D}_{ij}=-\hat{y}D$, ${\bf n}_i\times{\bf n}_j=-\hat{y}\sin(\theta_{m+1,n}-\theta_{m,n})=-\hat{y}\sin(2\pi/N_p)$, and thereby the total DMI energy is $E_{\rm DMI} = -ND\sin(2\pi/N_p)$. For the total SNN exchange energy $E_{J2}$, we sum up terms like ${\bf n}_i\cdot{\bf n}_j=-\cos(\theta_{m\pm1,n}-\theta_{m,n})=\sigma\cos(2\pi/N_p)$, where $\sigma=1$ for positive exchange interaction (phases FM and $c$-AFM) and $\sigma=-1$ for negative exchange interaction (phase $p$-AFM). Therefore, for all three cases, $E_{J2}=-2|J_2|\cos(2\pi/N_p)$. Finally, the contribution of anisotropy to the total energy is given by $E_K=-K\sum_{m,n}\cos^2\theta_{m,n}$, which, by assuming that $N/N_p$ is an integer, leads to $E_K=-\frac{1}{2}NK$. The total energy of the cycloidal phases is then given by $E/N = -2|J_2|\cos(2\pi/N_p)-D\sin(2\pi/N_p)-K/2$. Minimizing the energy with respect to $N_p$ yields
\begin{equation}
N_p=\frac{2\pi}{\tan^{-1}(D/2|J_2|)}
\end{equation}
and 
\begin{equation}\label{eq.E_cycloidal}
\frac{E}{N} = -\sqrt{4|J_2|^2+D^2}-\frac{K}{2}.
\end{equation}

Similar expressions can be obtained for the case of dominating NN exchange interaction, leading to the well-known dependence of the cycloid period on the ratio $D/J$\cite{belemuk2017phase,buhrandt2013skyrmion}. In order to delineate the boundaries of the cycloidal phases, we compare Eq.~\eqref{eq.E_cycloidal} with other (non-cycloidal) candidates to ground-state configuration. First, we consider phase V (see Fig.~\ref{fig3}). This phase is dominated by DMI interactions, where all neighboring spins are orthogonal to each other, so that $E_{J2}=0$. The DMI and anisotropy contribution to the energy can be calculated trivially, resulting in a total energy per spin $E/N = -\sqrt{2}D-K/2$. The critical DMI value, calculated by comparing this result with Eq.~\eqref{eq.E_cycloidal}, is given by $D_c=2|J_2|$ [see solid black lines in Fig.~\ref{fig3}(A)], irrespective of the anisotropy parameter $K$. Below this value, the cycloidal phases are favored against the homogeneous configuration V.

As shown in Fig.~\ref{fig3}(C) and (D), the cycloidal phases also become unstable above a certain value of the anisotropy parameter. For positive $K$, the $c$-AFM phase (III) is potentially favored for $J_2>0$, while for $J_2<0$, the $p$-AFM (VII) becomes the potential candidate. Their energy per spin is simply $E/N=-2|J_2| - K$. Comparing with Eq.~\eqref{eq.E_cycloidal}, we get $K_c= 2\sqrt{4|J_2|^2+D^2}-4|J_2|$ (see solid black lines in Fig.~\ref{fig3}(C) and (D)). 

For $K\leq0$, the cycloids decay into phases I and II, which are significantly influenced by the dipole-dipole interaction. For those cases, the demagnetizing energy can be approximated by an in-plane contribution to the effective anisotropy\cite{coey2010magnetism} $K_\text{eff}=K-K_\text{ddi}$, and the critical curves for $K\leq0$ are simply shifted by a factor of $K_\text{ddi}$ (see dashed and dashed-dotted lines in Fig.~\ref{fig3}(C) and (D)). The same curves can be plotted for the case $K=0$ as shown in Fig.~\ref{fig3}(A), where we used $K_\text{ddi}= 3.8D_\text{ddi}$ and $2D_\text{ddi}$ for the phase boundaries II-VI and I-IV respectively. 

\begin{figure}[t]
\centering
\includegraphics[width=0.7\linewidth]{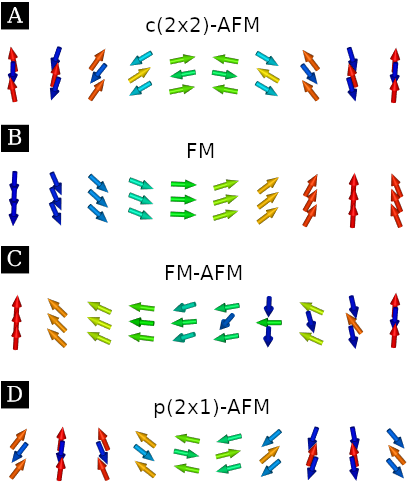}
\caption{N\'eel Domain wall structure for the $c(2$x$2)$-AFM state (A), FM state (B), between FM and AFM states (C) and $p(2$x$1)$-AFM state (D).  }
\label{fig4}
\end{figure}

\subsubsection{Domain walls and magnetic skyrmions}\label{s3.4}

Domain walls (DW) and magnetic skyrmions are promising candidates for technological applications, especially for spin-based information processing and memory devices. In this section we look for such magnetic configurations in the limit of suppressed NN exchange coupling. Fig.~\ref{fig4} summarizes different types of domain walls obtained in our simulations. For $J_2>0$ the well-known $c$-AFM DW [Fig.~\ref{fig4}(A)] and FM DW [Fig.~\ref{fig4}(B)] are found. However, since the FM and $c$-AFM states can coexist in this region of the phase diagram, a DW between the two phases is necessary to stabilize the spin system, as shown in Fig.~\ref{fig4}(C). In this case, only one sublattice of the $c$-AFM structure is rotated to form the FM state. For $J_2<0$ the $p$-AFM DW [Fig.~\ref{fig4}(D)] is the only one found. Due to such possibilities for different types of DW one also expects a variety of magnetic skyrmions in this system. Fig.~\ref{fig3}(A) shows an example of FM and $c$-AFM skyrmions coexisting for $J_2>0$ (see also supplemental figure Fig.~S1(B) for a better visualization of the coexisting FM and $c$-AFM skyrmions), and Fig.~\ref{fig3}(B) shows the $p$-AFM skyrmion obtained for $J_2<0$ - which to our knowledge is the first such skyrmion reported in literature. One should note that the FM-AFM DW [as shown in Fig.~\ref{fig4}(C)] can not form a stable topological object by enclosing itself (contrary to its FM or AFM counterparts, since in this case only half of the spins would rotate across the domain wall and the resultant net topological charge is non-integer).
\begin{figure}[t]
\centering
\includegraphics[width=\linewidth]{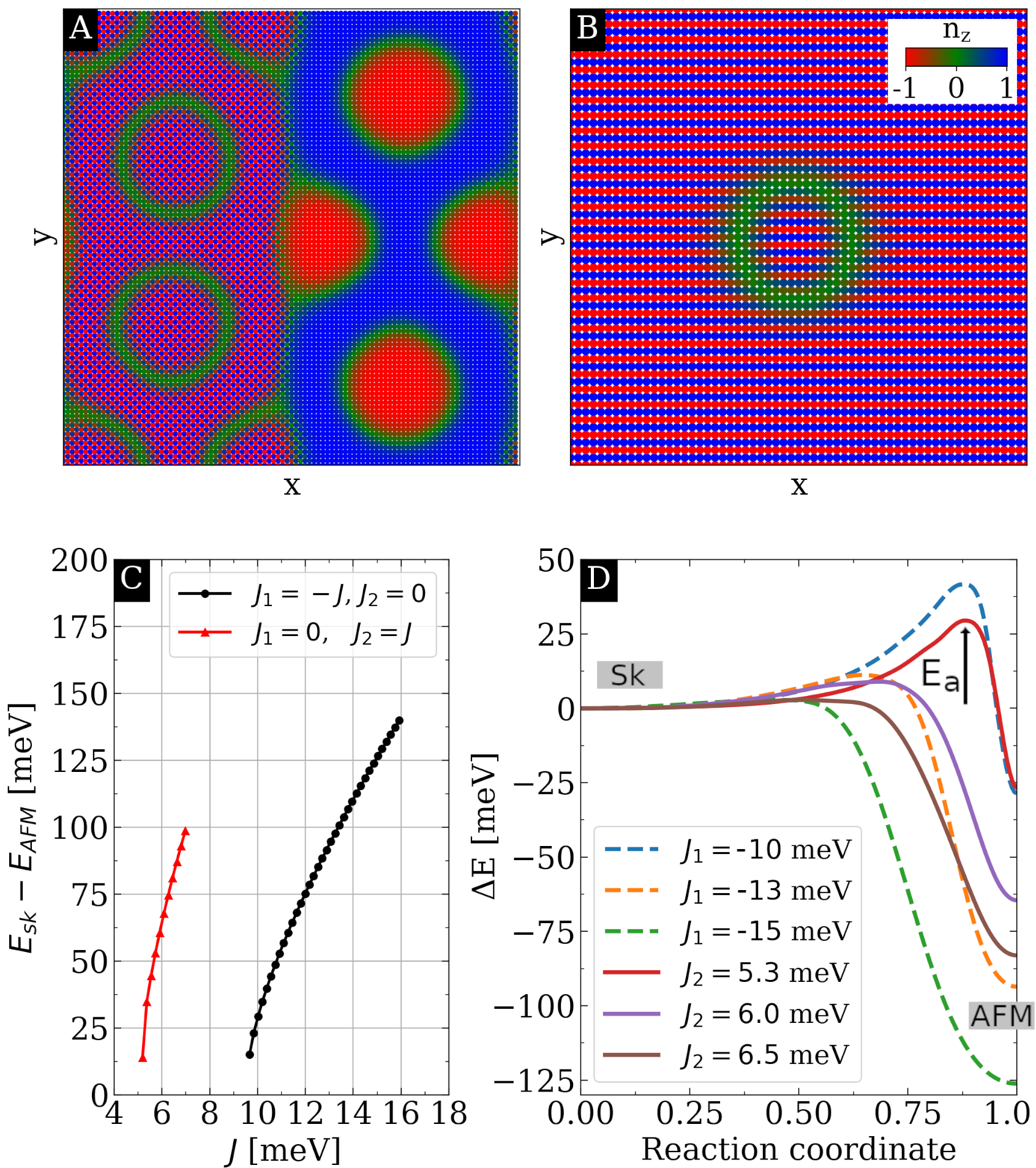}
\caption{(A) State with coexisting FM and $c$-AFM skyrmions, obtained for $J_2>0$. (B) The $p$-AFM skyrmion, obtained for $J_2<0$. (C) Skyrmion energy as a function of exchange coupling for (i) $J_1=0$, $J_2=J$, and (ii) $J_1=-J$, $J_2=0$, both with $K=3$~meV and $D=5$~meV. (D) Minimal-energy-path calculations for the isotropic collapse of the $c$-AFM skyrmion in both cases (i) and (ii). Here, the reaction coordinate defines the normalized (geodesic) displacement along the formation path and the activation energy $E_a$ is defined by the highest-energy point along the path.}
\label{fig5}
\end{figure}

Now, let us compare the influence of both NN and SNN exchange on the skyrmion stability. For that purpose we analyze the energy of an isolated $c$-AFM skyrmion, which can be stabilized by both $J_1$ and $J_2$. In Fig.~\ref{fig5}(C) we show the $c$-AFM skyrmion energy calculated for (i) $J_1=0$, $J_2=J$, and (ii) $J_1=-J$, $J_2=0$, where we fixed the parameters $K=3$~meV and $D=5$~meV in such a way that the skyrmion can be stabilized for $|J_2|$ similar to that obtained in Refs.~\onlinecite{ferriani2007magnetic,tan2019tunable}, where $|J_1|\approx0$. Notice that in case (i) one can stabilize skyrmions with equivalent energy but for smaller values of $J$ when compared to the case (ii), which indicates that SNN exchange can dominate the skyrmion energy over small values of $J_1$. Fig.~\ref{fig5}(D) shows the minimal-energy-path calculations for the isotropic collapse of the $c$-AFM skyrmion in both cases (i) and (ii), where the activation energy, $E_a$, is calculated by the geodesic nudged elastic band method~\cite{bessarab2015method,stosic2017paths} with the help of a climbing image method~\cite{henkelman2000improved}, both implemented in the simulation package \emph{Spirit}, allowing an accurate determination of the highest-energy saddle point along the minimal energy path connecting the two states. Notice that, for both cases (i) and (ii), the activation energies have the same order of magnitude, which indicates that the skyrmions stabilized by dominating SNN exchange interaction present similar stability, and consequently similar lifetime\cite{bessarab2019stability,stosic2017paths}, to those stabilized by dominating NN exchange coupling. Finally, the stability of the $p$-AFM skyrmion was found to be the same as the stability of the $c$-AFM skyrmion, as long as the same values of $|J_2|$ are considered. 

\subsection{Suppressed nearest-neighbor exchange in the honeycomb lattice}\label{ss3C}

\begin{figure}[t]
\centering
\includegraphics[width=6cm]{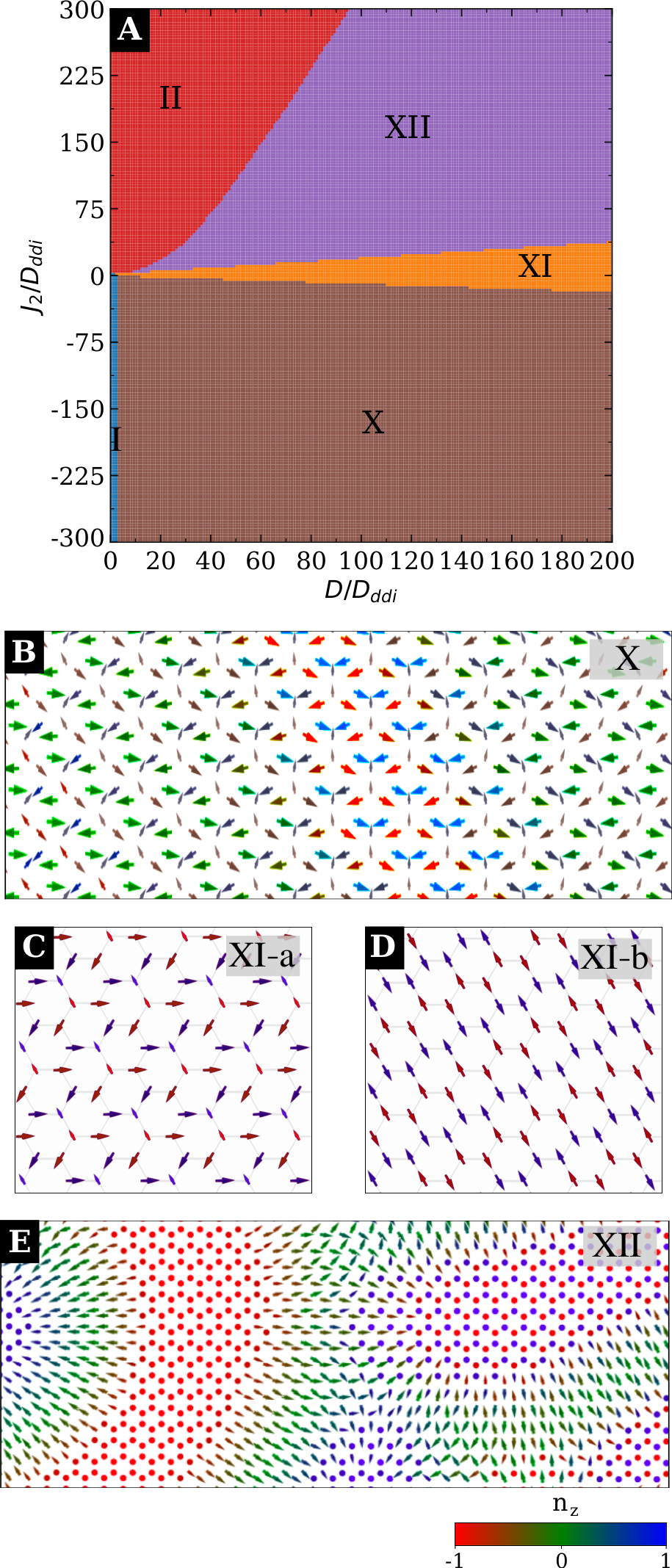}
\caption{(A) Phase diagram for the honeycomb lattice for varied second-nearest-neighbor exchange $J_2$ and DMI strength $D$, for vanishing nearest-neighbor exchange and $K=0$. (B-E) The spin textures corresponding to the characteristic magnetic phases in panel (A).}
\label{fig6}
\end{figure}

\begin{figure}[t]
\centering
\includegraphics[width=\linewidth]{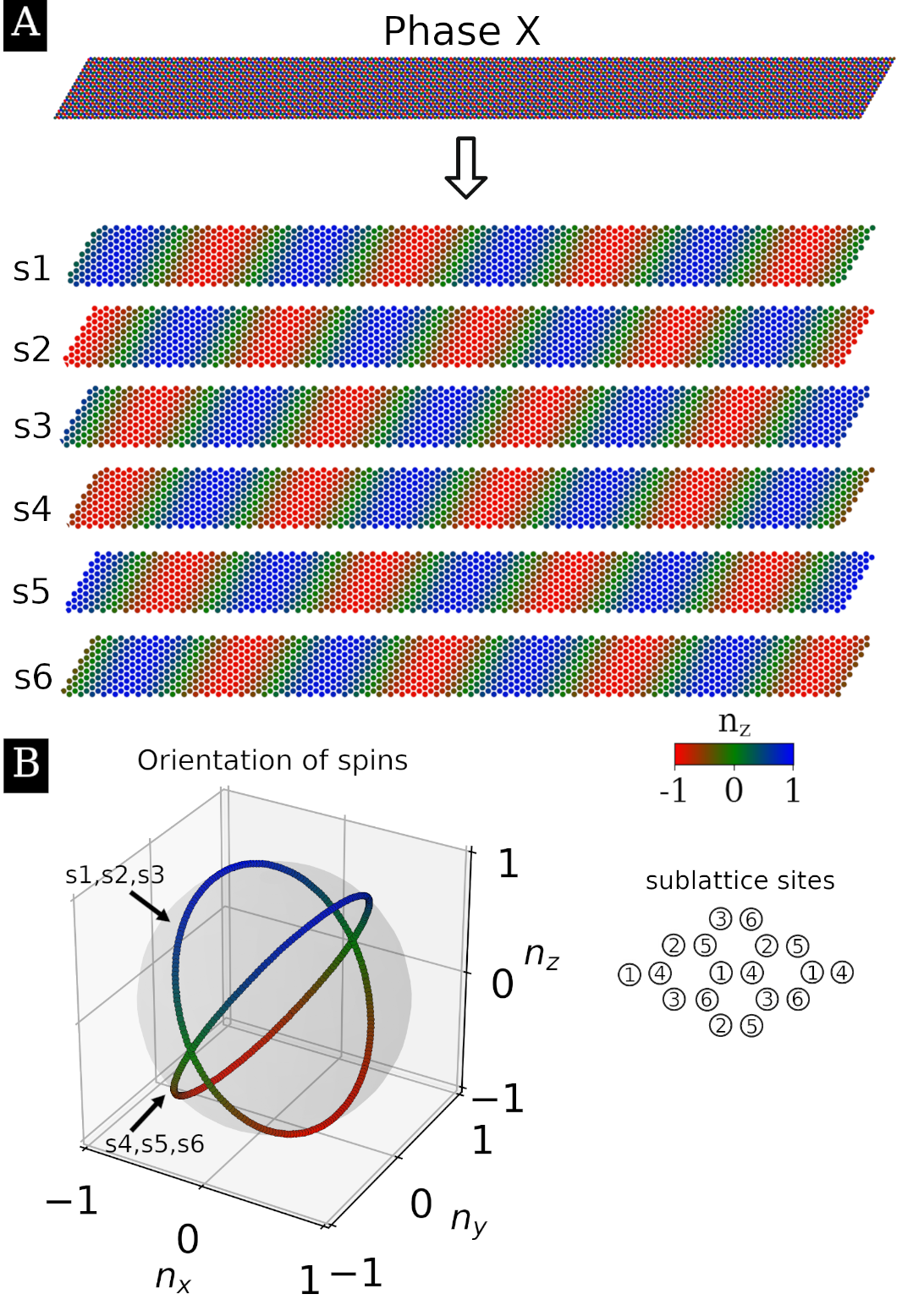}
\caption{(A) Decomposition of phase X (obtained in the phase diagram for the honeycomb lattice for $J_1=0$, $J_2<0$ and $D>0$) into six spin-sublattices (s1-s6). (B) Orientation of spins around the unit sphere, demonstrating the sublattice-specific cycloidal behavior.}
\label{fig7}
\end{figure}

\begin{figure}[t]
\centering
\includegraphics[width=\linewidth]{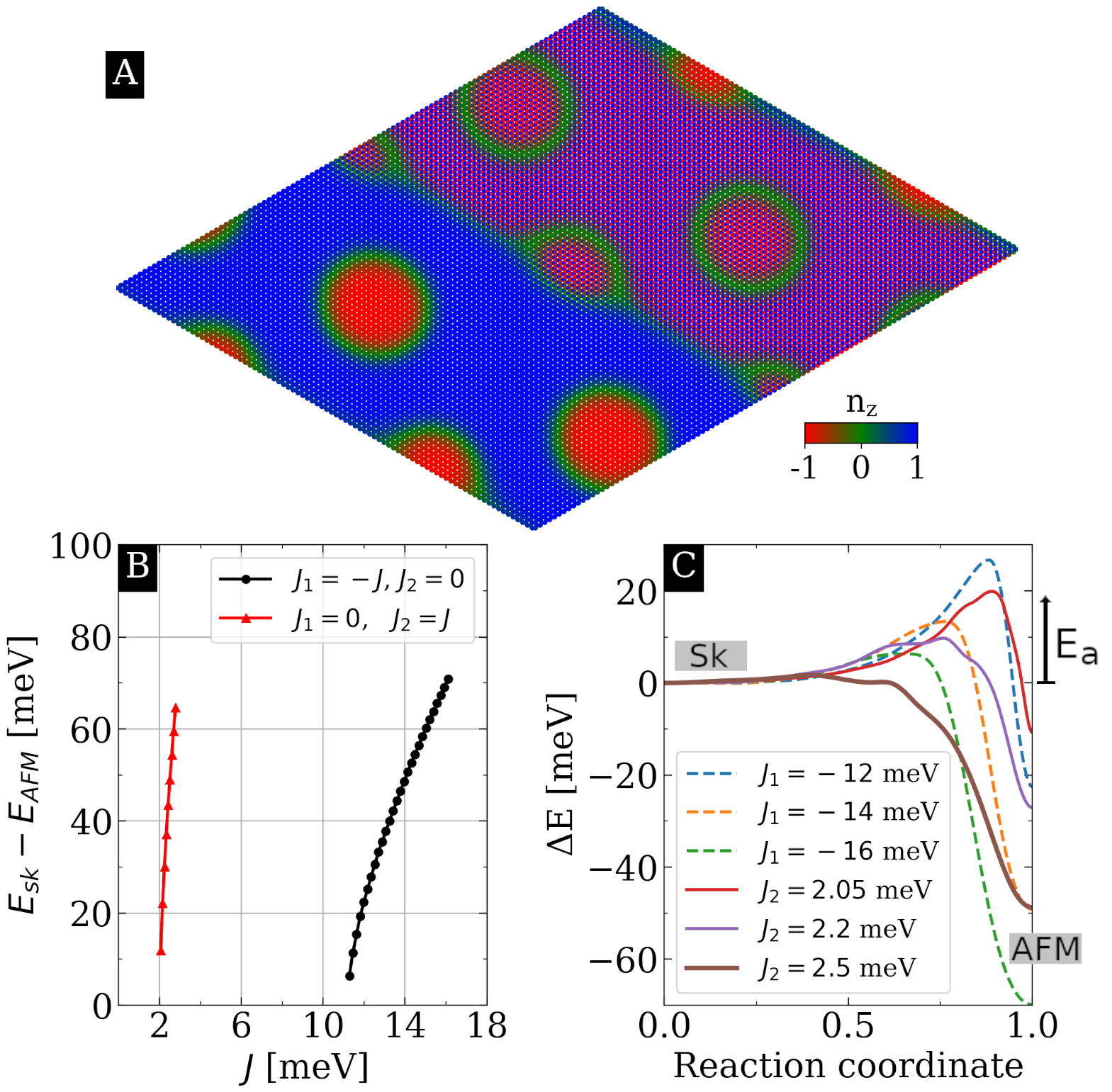}
\caption{(A) State with coexisting FM and AFM skyrmions, obtained for $J_2>0$ in the honeycomb symmetry. (B) Skyrmion energy as a function of exchange coupling for (i) $J_1=0$, $J_2=J$, and (ii) $J_1=-J$, $J_2=0$, both with $K=3$~meV and $D=5$~meV. (D) Minimal-energy-path calculations for the isotropic collapse of the AFM skyrmion in both cases (i) and (ii). Here, the reaction coordinate defines the normalized (geodesic) displacement along the formation path and the activation energy $E_a$ is defined by the highest-energy point along the path.}
\label{fig8}
\end{figure}

The honeycomb lattice symmetry is representative of a variety of magnetic 2D materials, and is hence of particular recent interest. Notice that for a complete description of truly 2D magnetic materials (such as monolayer CrI$_3$) one should consider more magnetic interactions in the Hamiltonian, such as the Kitaev interaction \cite{xu2018interplay}. In the present consideration, for simplicity and clarity, but also in order to provide a fair comparison between square and honeycomb lattice symmetry, we do not include those additional magnetic interactions in our spin system. Fig.~\ref{fig6}(A) exhibits the phase diagram obtained for the honeycomb lattice, for zeroed NN exchange coupling ($J_1$) and anisotropy ($K$), in the presence of dipole-dipole interactions, and for varied SNN exchange ($J_2$) and DMI strength ($D$). Similarly to the square lattice, for $J_2>0$, increasing DMI favors FM and AFM spin cycloids (see Fig.~\ref{fig6}(E), and also Fig.~S1(C) for a better visualization of the coexisting FM and AFM states). For dominating DMI interaction all spins are forced orthogonal to each other, as shown in Fig.~\ref{fig6}(C) and Fig.~\ref{fig6}(D) (phase XI). Since in the honeycomb lattice the SNN bonds form a triangular lattice, which is frustrated under AFM coupling, the analogous forms of the $p$-AFM state and the $p$-AFM skyrmion found in the square lattice can not be obtained for $J_2<0$. Instead, for $J_2<0$ we obtained frustrated AFM spin cycloids (shown in Fig.~\ref{fig6}(B), but can be visualized better in supplementary Fig.~S2). In Fig.~\ref{fig7}(A) we show that this configuration can actually be decomposed into six sublattices, each of which exhibiting a cycloidal behavior - as seen in the orientation of spins shown in Fig.~\ref{fig7}(B). Due to the honeycomb symmetry of the lattice, the anti-parallel AFM coupling between the SNN spins can not be satisfied at all bonds and the system is therefore frustrated, causing this unique cycloidal spin state.

To conclude the comparison to our previous findings for the square lattice, in Fig.~\ref{fig8}(A) we show an example of FM and $c$-AFM skyrmions coexisting for $J_2>0$ in the honeycomb lattice symmetry (see also supplemental figure Fig.~S1(D) for a better visualization of the coexisting FM and AFM skyrmions). Similarly to the analysis made in the preceding section, we have calculated the energy of an isolated AFM skyrmion and performed minimal-energy-path calculations to evaluate the influence of both NN and SNN exchange to the skyrmion stability. As in the square lattice, both kinds of skyrmions have equivalent energies [see Fig.~\ref{fig8}(C)] but their stability falls in different ranges of the exchange intensity $|J|$, with the SNN-stabilized skyrmion occurring at considerably smaller values of $|J|$. Finally we note again that, contrary to the square-lattice case, skyrmions with a $p$-AFM background are not possible in the honeycomb lattice because, as discussed above, AFM ordering is frustrated on a triangular lattice of second-nearest neighbors. 

\section{Conclusions}\label{s5}      
The advent of monolayer materials over the last decade, freestanding or deposited, raises questions whether such systems can harbor unique magnetic properties and potential for technological applications. It is primarily the vast tunability of magnetic interactions in such magnetic monolayers that can lead to unexpected physical phenomena and magnetic phases. Our present work is a step in that direction, revealing the rich magnetic phase diagram for both square and honeycomb symmetries of a magnetic monolayer, in the limit of suppressed nearest-neighbor exchange interaction. With underlying expectation of degenerate ferromagnetic and antiferromagnetic states promoted by the absence of the nearest-neighbor exchange, the competition between the second-nearest-neighbor exchange, DMI, and dipolar interactions leads to several unique cycloidal, checkerboard, row-wise and spin-ice states, unattainable otherwise. Moreover, coexisting FM and AFM skyrmions are found, as well as novel types of chiral domain walls and skyrmions (such as the $p$-AFM ones). With several existing \textit{ab initio} predictions that exchange interactions in magnetic monolayers can be varied depending on the substrate\cite{ferriani2007magnetic,tan2019tunable, meyer2017dzyaloshinskii,hardrat2009complex}, the states found in our work will serve as a smoking gun for the experimental validation of such claims. Last but not least, the interactions in the recently realized 2D magnetic materials can also be broadly manipulated by e.g. strain engineering\cite{webster2018strain}, where our present results can provide basic expectations and understanding before the more detailed calculations (including additional anisotropic exchange couplings such as Kitaev one\cite{kitaev2006anyons}) are performed.

\section*{Acknowledgements}
This work was supported by the Research Foundation - Flanders (FWO-Vlaanderen) and Brazilian Agencies FACEPE (under grant No. APQ-0198-1.05/14), CAPES and CNPq.


\bibliography{references}

\end{document}